\documentclass[aps, amsmath,amssymb,
superscriptaddress, showpacs,prl,
preprint]{revtex4-1}

\setcitestyle{super}

\usepackage{graphicx}
\usepackage{multirow}
\usepackage{upgreek}
\usepackage{epsf}
\usepackage{epstopdf}
\usepackage{units}
\usepackage{wrapfig}
\usepackage{subfigure}

\newcommand{\e}{\mathrm{e}}

\begin{document}

\title{Multi-Pulse Laser Wakefield Acceleration: A New Route to Efficient, High-Repetition-Rate Plasma Accelerators and High Flux Radiation Sources}
\author{S. M. Hooker}
\email{simon.hooker@physics.ox.ac.uk.}
\affiliation{John Adams Institute for Accelerator Science, Department of Physics, University of Oxford, OX1 3PU, United Kingdom}

\author{R. Bartolini}
\affiliation{John Adams Institute for Accelerator Science, Department of Physics, University of Oxford, OX1 3PU, United Kingdom}
\affiliation{Diamond Light Source, Oxfordshire, OX11 0DE, United Kingdom}

\author{S. P. D. Mangles}
\affiliation{John Adams Institute for Accelerator Science, The Blackett Laboratory, Imperial College London, London SW7 2AZ, United Kingdom}

\author{A. T\"unnermann}
\affiliation{Institute of Applied Physics, Abbe Center of Photonics, Friedrich-Schiller-Universit\"at Jena, 07743 Jena, Germany}
\affiliation{Fraunhofer Institute for Applied Optics and Precision Engineering, Jena, Germany}

\author{L. Corner}
\affiliation{John Adams Institute for Accelerator Science, Department of Physics, University of Oxford, OX1 3PU, United Kingdom}

\author{J. Limpert}
\affiliation{Institute of Applied Physics, Abbe Center of Photonics, Friedrich-Schiller-Universit\"at Jena, 07743 Jena, Germany}
\affiliation{Fraunhofer Institute for Applied Optics and Precision Engineering, Jena, Germany}

\author{A. Seryi}
\affiliation{John Adams Institute for Accelerator Science, Department of Physics, University of Oxford, OX1 3PU, United Kingdom}

\author{R. Walczak}
\affiliation{John Adams Institute for Accelerator Science, Department of Physics, University of Oxford, OX1 3PU, United Kingdom}

\date{\today}

\begin{abstract}
Laser-driven plasma accelerators can generate accelerating gradients three orders of magnitude larger than radio-frequency accelerators and have achieved beam energies above \unit[1]{GeV} in centimetre long stages. However, the pulse repetition rate and wall-plug efficiency of plasma accelerators is limited by the driving laser to less than approximately \unit[1]{Hz} and 0.1\% respectively. Here we investigate the prospects for exciting the plasma wave with trains of low-energy laser pulses rather than a single high-energy pulse. Resonantly exciting the wakefield in this way would enable the use of different technologies, such as fibre or thin-disc lasers, which are able to operate at multi-kilohertz pulse repetition rates and with wall-plug efficiencies two orders of magnitude higher than current laser systems. We outline the parameters of efficient, GeV-scale, 10-kHz plasma accelerators and show that they could drive compact X-ray sources with \emph{average} photon fluxes comparable to those of third-generation light source but with significantly improved temporal resolution. Likewise FEL operation could be driven with comparable peak power but with significantly larger repetition rates than extant FELs.
\end{abstract}


\maketitle

Many scientific fields use ultrafast pulses of radiation to probe dynamical processes. Much of this work is performed with synchrotrons and the new generation of x-ray free-electron lasers (FELs), both of which are powered by energetic electron beams.\cite{Emma:2010} These facilities have played a pivotal role in driving progress in  the physical, biological, and medical sciences, and this will continue to be the case. However, their high cost and large scale --- both primarily determined by that of the radio-frequency electron accelerators which drive them --- necessarily limits access, and therefore restricts the amount, and potentially the nature, of the research that can be done.

Laser-driven plasma accelerators can generate electron beams with energies of a few GeV --- comparable to that  used in synchrotrons and FELs --- but in accelerator stages only a few centimetres long.\cite{Leemans:2006, Kneip:2009, Wang:2013} Plasma accelerators could therefore drive very compact sources of femtosecond-duration pulses of radiation ranging from THz frequencies to the X-ray range which, importantly, are naturally synchronized to the femtosecond driving laser pulse. Indeed laser-accelerated electron beams have already been used to generate incoherent undulator radiation with photon energies up to about \unit[100]{eV},\cite{Fuchs:2009} and incoherent betatron radiation in the \unit[10]{keV} range.\cite{Kneip:2010}

However, the application of plasma accelerators to driving useful radiation sources (and, in the longer term, particle colliders) is prevented by the low repetition rate $f_\mathrm{rep}$ (of order \unit[1]{Hz}) and wall-plug efficiency (less than 0.1\%) of the driving laser. In this paper we investigate the prospects for multi-pulse laser wakefield acceleration (MP-LWFA), in which the wake is excited by a \emph{train} of low-energy laser pulses rather than by a single high-energy pulse. Moving the problem of energy storage from the laser medium to the plasma would allow the use of novel laser technologies, such as fibre or thin-disc lasers, which can operate with $f_\mathrm{rep}$ in the multi-kilohertz range and with high wall-plug efficiency.

We note that MP-LWFA was studied theoretically\cite{Nakajima:1992, Berezhiani:1992vw, Umstadter:1994, Johnson:1994, Dalla:1994b, Bonnaud:1994uo, Umstadter:1995, Kalinnikova:2008} in the 1990s, but has yet to be demonstrated. Driving plasma wakefields with a train of \emph{particle} bunches or a modulated particle bunch has also been investigated theoretically,\cite{Kallos:2008, Caldwell:2011} and generation of an acceleration gradient of $\unit[22]{MV m^{-1}}$ by a train of 7 electron bunches has been demonstrated experimentally.\cite{Caldwell:2008}

In this  paper we build on our earlier work\cite{Corner:2012} on the prospects of driving LWFAs with trains of laser pulses, by considering the potential issues of this approach and showing that MP-LWFAs could drive: (i) compact incoherent X-ray sources with \emph{average} photon flux comparable that of time-resolved third-generation radiation facilities (and with significantly superior temporal resolution); and (ii) FELs with comparable peak power to existing machines, but with significantly higher repetition rates.

\section{Multi-pulse laser wakefield acceleration}
In a laser wakefield accelerator (LWFA) a single laser pulse, with a peak intensity of order $\unit[10^{18}]{W cm^{-2}}$, propagates through a plasma and excites a density wave via the ponderomotive force, which acts to expel plasma electrons from the region of the laser pulse. As described in recent reviews,\cite{Esarey:2009, Hooker:2013} the electric fields developed within the plasma wave are of the order of $\unit[100]{GV m^{-1}}$ --- some three orders of magnitude greater than possible with a conventional accelerator.

\begin{figure}
\centering
\subfigure[]{\includegraphics[width=0.45\linewidth]{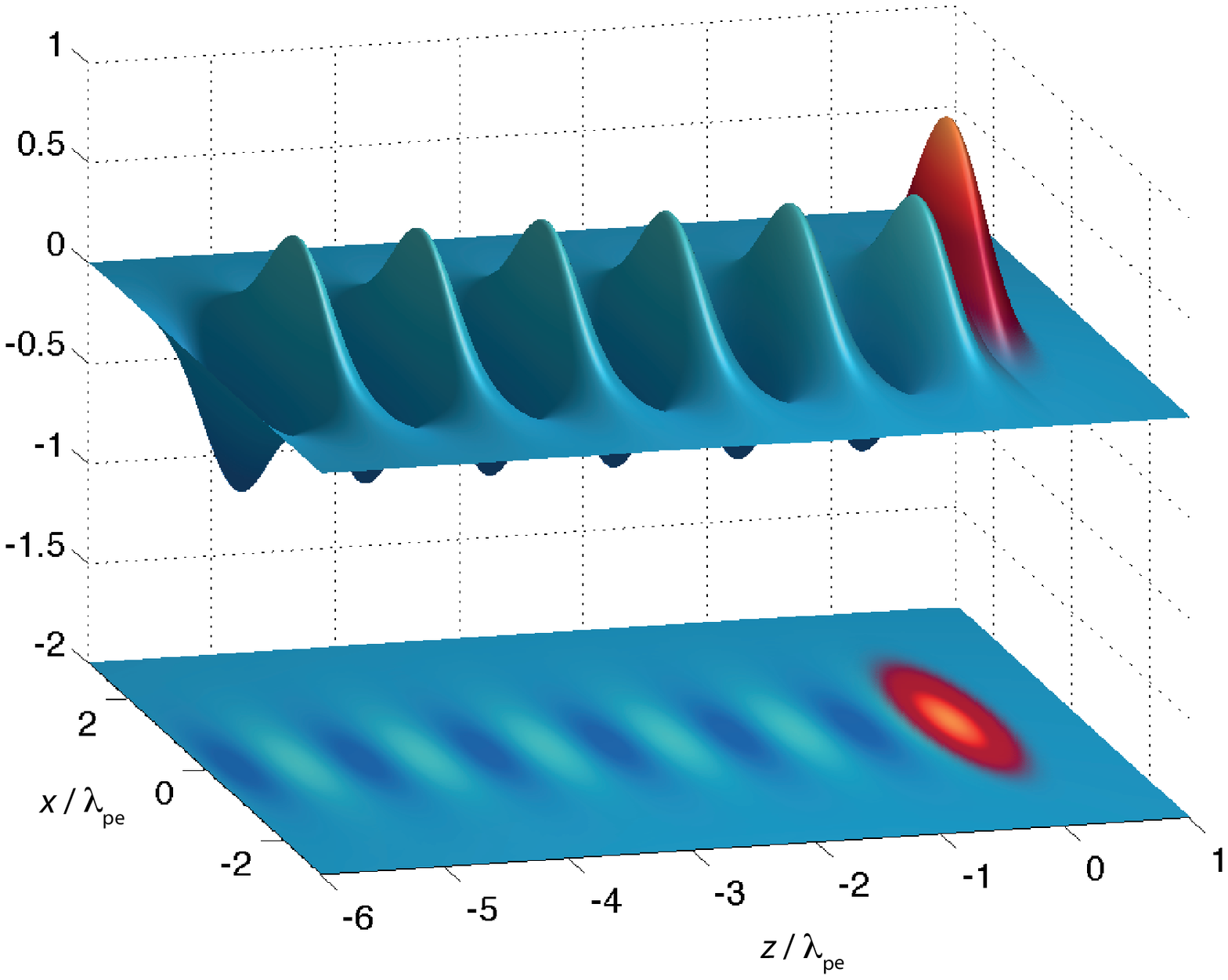}}
\subfigure[]{\includegraphics[width=0.45\linewidth]{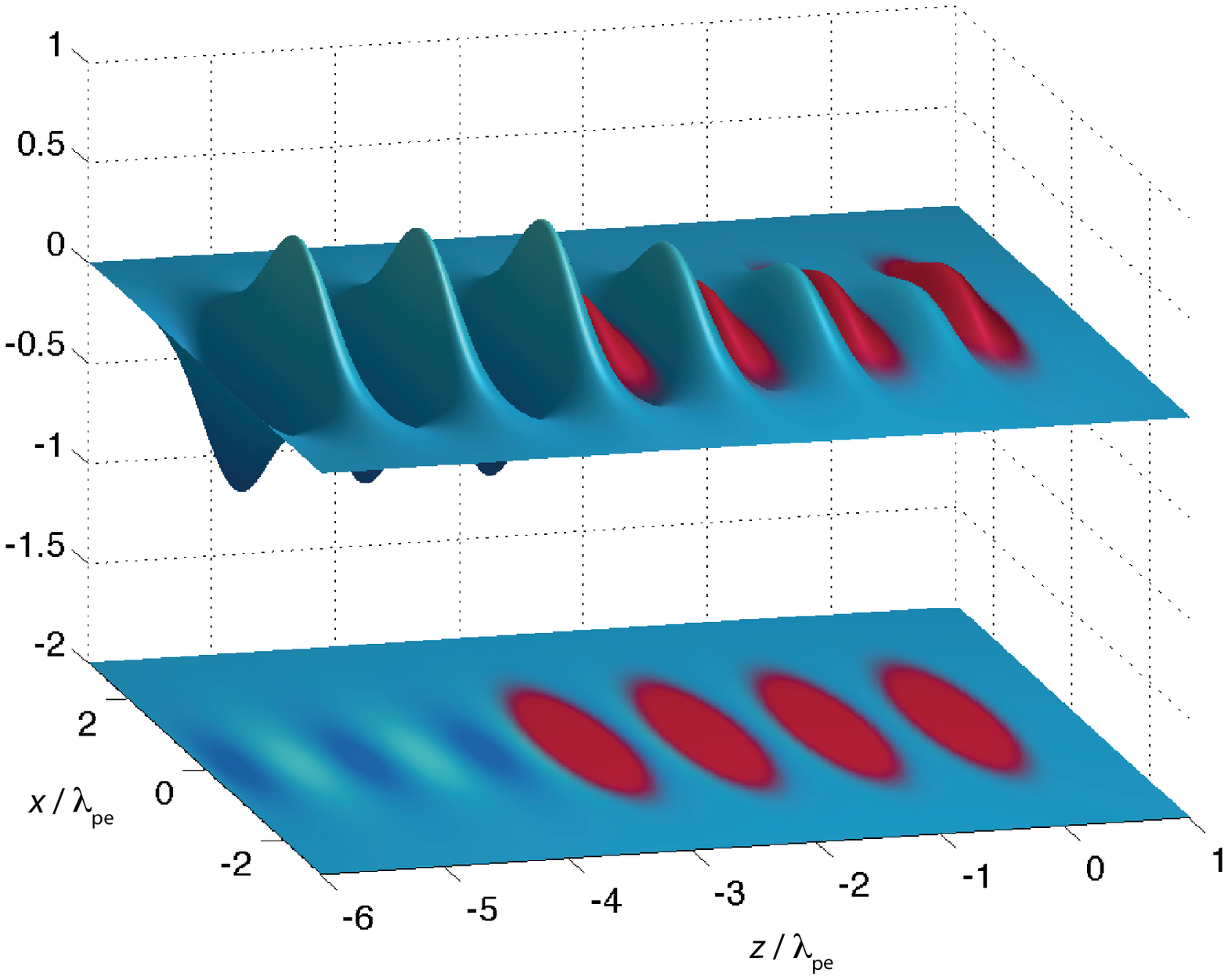}}
\caption{Comparison of the LWFA and MP-LWFA schemes, showing the electron density of the plasma (blue) and laser pulse intensity (red) for the case of: (a) LWFA, in which the wakefield is driven by a single, high-energy laser pulse; (b) MP-LWFA, in which a train of low-energy laser pulses excites the plasma wave. The laser pulses (red) propagate along the $z$-axis, towards positive $z$, and their intensities are given in terms of $a_0^2$, where $a_0$ is the normalized vector potential. The electron density is given in terms of $\Delta n_\e / n_\e$, where $\Delta n_\e$ is the increase in the electron density from its mean value $n_\e$. The longitudinal ($z$) and transverse ($x$) spatial co-ordinates are normalized by the plasma wavelength $\lambda_\mathrm{pe} = 2 \pi c /\omega_\mathrm{pe}$.}\label{Fig:Concepts}
\end{figure}

In MP-LWFA a train of low-energy laser pulses, rather than a single high-energy pulse, is used to excite the plasma wave. The plasma wakefields excited by the pulses will add coherently --- so that the amplitude of the wakefield increases with each additional pulse --- if the pulses are spaced by the plasma period $T_\mathrm{pe} = 2 \pi / \omega_\mathrm{pe}$, where  $\omega_\mathrm{pe} = (n_\mathrm{e} e^2 / m_\mathrm{e} \epsilon_0)^{1/2}$ and $n_\mathrm{e}$ is the mean electron density. Excitation of a plasma wakefield by single and multiple laser pulses is illustrated schematically in Fig.\ \ref{Fig:Concepts}.

We note that driving a wakefield with a train of $N$ pulses has several potential advantages over LWFA. First, the driving laser energy is spread over the duration of the pulse train, reducing by a factor of $N$ the peak laser intensity to which the optical components are exposed. Importantly, the reduced intensity would enable the use of an optic of small-diameter $D$ and short focal length $f \propto D \propto 1/\sqrt{N}$ to couple the pulse train into the plasma; this would considerably reduce the space required for staged plasma accelerators\cite{Schroeder:2010} and potentially avoid the need for plasma mirrors.  Second, MP-LWFA offers scope for additional control by adjusting the spacing, centre frequency, energy, and duration of the pulses in the train; this flexibility is likely to be important for developing advanced plasma accelerators which are optimized for efficiency, which are stable against longitudinal variations of the plasma density\cite{Lindberg:2006gz} and/or which exploit longitudinal tapering to achieve acceleration beyond the dephasing length.\cite{Bulanov:1998,Sprangle:2001, Rittershofer:2010} Third, it has been shown that large amplitude (nonlinear) wakefields can be excited more efficiently, and with reduced plasma instabilities, by a train of pulses than with a single pulse of the same total energy.\cite{Berezhiani:1992vw, Umstadter:1994, Johnson:1994, Dalla:1994b, Bonnaud:1994uo, Umstadter:1995}

\subsection{Numerical modelling of a MP-LWFA}

In order to develop a more quantitative understanding of the operation of a MP-LWFA we have investigated the plasma wave that could be driven by a pulse train with parameters similar to that which could be produced by a state-of-the-art, high-repetition fibre laser system, as described below. Our calculations were performed  using weakly relativistic electron fluid equations \cite{Miano:1990wf} and the particle-in-cell code \textsc{osiris}.\cite{OSIRIS} Each laser pulse in the train was assumed to have an energy of \unit[10]{mJ},  a pulse duration of $\tau_\mathrm{FWHM} = \unit[100]{fs}$, and a centre wavelength of $\lambda = \unit[1]{\mu m}$; the pulses were taken to be focused to a  spot size of $w_0 = \unit[40]{\mu m}$, corresponding to a peak normalized vector potential of $a_0 = 0.052$ and spaced by the plasma period. Calculations were performed for a plasma density of $n_e = \unit[1.75 \times 10^{17}]{cm^{-3}}$, which maximises the electron energy gain produced by this pulse train.

\begin{figure}[tb]
\centering
\includegraphics[width=0.8\linewidth]{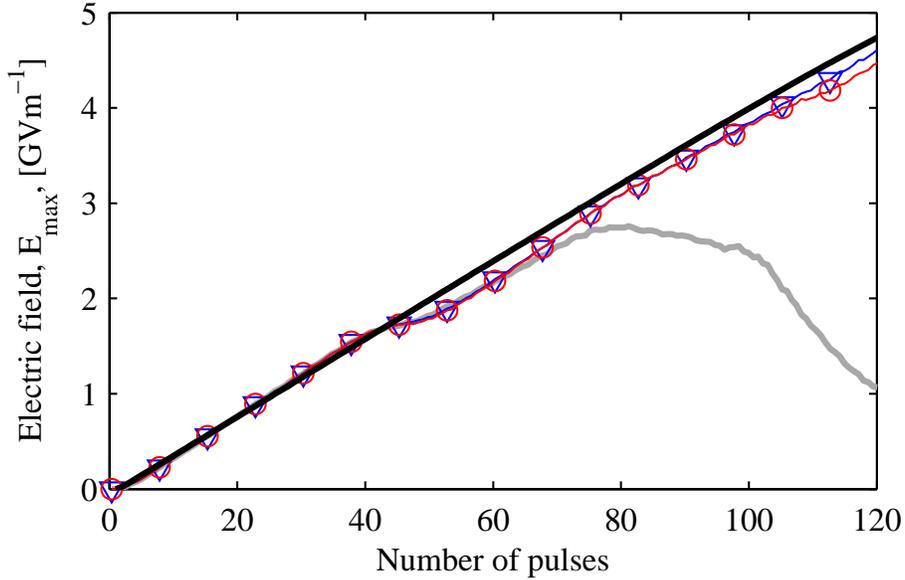}
\caption{Calculated maximum accelerating field $E_\mathrm{acc}^\mathrm{max}$ within the plasma wave as a function of the number of pulses in the pulse train described in the main text. The solid black line shows the results of fluid simulations, and the results of PIC simulations are shown for the case of stationary (blue) and mobile (red) Xe$^{8+}$ ions, and for mobile hydrogen ions (grey).}
\label{Fig:Fluid_sims}
\end{figure}

Figure \ref{Fig:Fluid_sims} shows the growth predicted by fluid simulations of the maximum axial accelerating field $E_\mathrm{acc}^\mathrm{max}$ as a function of the number of the pulse within the train; it can be seen that $E_\mathrm{acc}^\mathrm{max}$ grows linearly.  The amplitude of the plasma density wave (not shown) grows in similar way and reaches a relative value of $\Delta n_e / n_e = 22\%$ on axis. This corresponds to only a mildly nonlinear wakefield and hence there was no need to adjust the spacing between laser pulses to counter relativistic detuning. 

Our fluid simulations do not include ion motion, which is expected to be significant for pulse trains with a total duration which is not short compared to the ion plasma period $T_{pi} = 2\pi / \Omega_{pi}$ where  $\Omega_{pi} = \sqrt{ Z^2 n_i e^2 /(m_i\epsilon_0)}$.   To understand the effect of ion motion we performed PIC simulations for neutral plasmas comprising electrons and: (i) pre-ionized hydrogen ions; (ii) immobile pre-ionized Xe$^{8+}$ ions; and (iii) mobile pre-ionized Xe$^{8+}$ ions. The PIC simulations of Fig.\ \ref{Fig:Fluid_sims} show that ion motion will indeed limit the total number of pulses which can usefully be used in a MP-LWFA. For the chosen parameters the useful number of pulses is  $N \approx 120$, at which point the plasma wave accelerating field has reached $E_\mathrm{acc}^\mathrm{max} = \unit[4.7]{GV/m}$. For an accelerator with a length equal to half the dephasing length, $L_\mathrm{acc} = L_\mathrm{d} / 2 = \unit[260]{mm}$ --- corresponding  to acceleration in the accelerating \emph{and} focusing phases of the plasma wave --- the energy gain is $W_\mathrm{max} = (2 / \pi) E_\mathrm{acc}^\mathrm{max} L_\mathrm{acc} = \unit[0.75]{GeV}$.

\section{Lasers for MP-LWFA}\label{Sec:Lasers}
State-of-the-art femtosecond solid-state lasers used for particle acceleration are joule-class titanium-sapphire chirped-pulse amplification (CPA) systems. \cite{Leemans:2006} Lasers of this type can easily generate the high peak intensities required for driving plasma accelerators, but they can only operate at a relatively low average power due to their poor thermo-optical properties and low efficiency (typically much less than 1\%). In contrast, innovative diode-pumped solid-state lasers such as slab, thin-disc and fibre lasers are able to provide ultra-short laser pulses with \emph{average} power\cite{Tunnermann:2010} above \unit[1]{kW}, but \emph{peak} powers of only a few GW.

\begin{figure}[tb]
\centering
\includegraphics[width=0.8\linewidth]{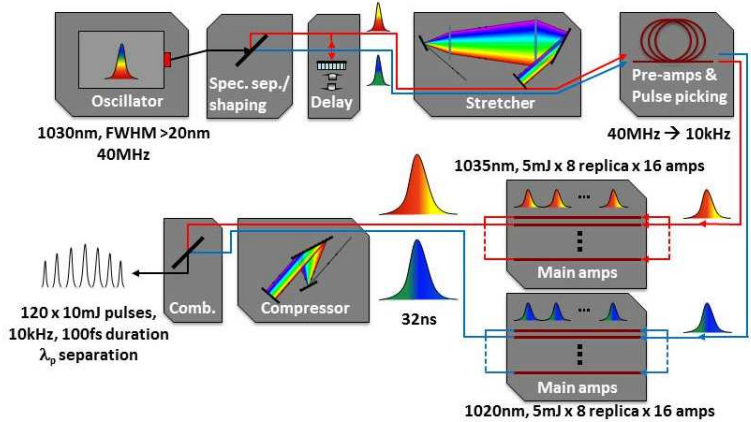}
\caption{Schematic setup of a fibre-based laser system capable of producing a train of 120 \unit[10]{mJ} pulses at a repetition rate of  \unit[10]{kHz}.}
\label{Fig:Laser_design}
\end{figure}

Recently, it has been proposed that the demanding requirements of LWFAs operating at high $f_\mathrm{rep}$ could be met by coherently combining the output of many lasers.\cite{Morou:2013} This approach involves spatially separated amplification of the pulses, which distributes the challenges imposed on the laser gain medium from one emitter to many and, in effect, operates as an amplifying interferometer. This scheme has been extensively investigated using optical fibres as the gain media, and performance beyond that possible from a single fibre amplifier has been demonstrated.\cite{Klenke:2013} This idea can be further extended into the temporal domain by splitting each pulse into a short train, which is amplified and then temporally recombined into a single high-energy pulse --- an approach known as divided-pulse-amplification (DPA).\cite{Zhou:2007} It is anticipated that a combination of spatial and temporal multiplexing will allow the production of joule-level pulses with $f_\mathrm{rep}$ in the kHz range.

Multi-pulse LWFAs could be driven by a related approach, but with the additional requirement of the generation of a train of 100-200 pulses. Pulse trains can be generated using beam splitters and delay lines,\cite{Siders:1998} or by employing spectral-shaping; however, these methods typically introduce losses exceeding 50\%, or have challenging alignment requirements. Here we outline one method for generating a pulse train suitable for MP-LWFA based on interference of two spectrally-separated pulses in a variation of the well-known plasma beat-wave accelerator.\cite{Esarey:2009, Rosenbluth:1972, Amiranoff:1992} In this approach broadband pulses are amplified in two CPA systems and partially compressed to the duration required of the entire pulse train. If the pulses have a constant angular frequency difference $\Delta \omega$, then when combined the temporal profile will correspond to that of the stretched pulses modulated by a cosine-squared function of period $\Delta T = 2 \pi / \Delta \omega$. We note that it may also be possible to generate a train of pulses with optimized parameters by phase-only filtering; this is compatible with CPA since it maintains a broad-band spectrum with minimal amplitude modulation.\cite{Weiner:1990}

Figure \ref{Fig:Laser_design} shows a conceptual design of a table-top laser system suitable for driving a MP-LWFA. Pulses from a femtosecond oscillator are separated into two spectral branches with centre wavelengths of \unit[1020]{nm} and \unit[1035]{nm}. Each spectral branch is amplified as follows: the pulse is stretched to a few nanoseconds duration; the repetition rate is reduced to \unit[10]{kHz} and the pulse is pre-amplified; the pulse is then amplified in a 16-amplifier DPA scheme in which the pulse is split into a train of 8 replicas; following temporal recombination, the pulse is partially recompressed. The two spectral branches are then combined to form the MP-LWFA pulse train of $N = 120$, \unit[10]{mJ}, \unit[100]{fs} pulses at $f_\mathrm{rep} = \unit[10]{kHz}$. 

An important feature of this approach is that it is possible to manipulate the spectral phases of the two stretched pulses so that the two spectral branches, when partially compressed, have a non-constant angular   frequency difference $\Delta \omega (t)$. Combining these branches will then generate a pulse train with controllable, non-uniform pulse spacing. The additional control enabled by phase manipulation of two spectrally-broad spectral branches will allow resonance to be maintained for nonlinear plasma wake-fields, as discussed by previous authors.\cite{Deutsch:1991} It also offers scope for achieving auto-resonance by down-sweeping the beat frequency across the local plasma frequency, as discussed by Lindberg et al.;\cite{Lindberg:2006gz} this could provide stability against variations in the local plasma density, or allow MP-LWFAs to achieve acceleration beyond the de-phasing length by employing tapered plasmas.\cite{Bulanov:1998,Sprangle:2001, Rittershofer:2010}

\section{Radiation sources driven by a kHz MP-LWFA}
\label{Sec:Radiation}

The high pulse repetition rate which could be achieved with MP-LWFAs would make them ideal drivers of femtosecond radiation sources providing high mean photon flux as well as exceptionally high peak brightness. Here we consider radiation generation in undulators and from the betatron motion of the transverse oscillation of the electrons as they accelerate in the plasma wakefield.  We note that radiation generated by other methods, such as Thomson and Compton scattering, \cite{Phuoc:2012cg, Powers:2013bx} would also benefit from the high values of $f_\mathrm{rep}$ enabled by MP-LWFA. In the calculations below we assume electron beam parameters consistent with existing LWFA performance\cite{Hooker:2013} and the fluid and PIC simulations given above: an electron energy of \unit[0.75]{GeV};  1\% relative energy spread; a bunch charge of \unit[50]{pC}; a normalized transverse emittance of $\unit[0.1 \pi]{\mu m}$, and a Gaussian temporal profile of \unit[5]{fs} FWHM.

\subsection{Betatron radiation}

As outlined in the Methods section, the radiation generated by betatron motion within the MP-LWFA can be estimated using standard theory,\cite{Esarey:2002hk} noting that for the parameters we have considered the MP-LWFA will not operate in the ``blow-out'' regime as is usually assumed. For efficient betatron radiation production it will be necessary to operate closer to the non-linear regime which can be achieved by using tighter focusing or higher plasma densities.

\begin{figure}[tb]
\centering
\includegraphics[width = 8.9cm]{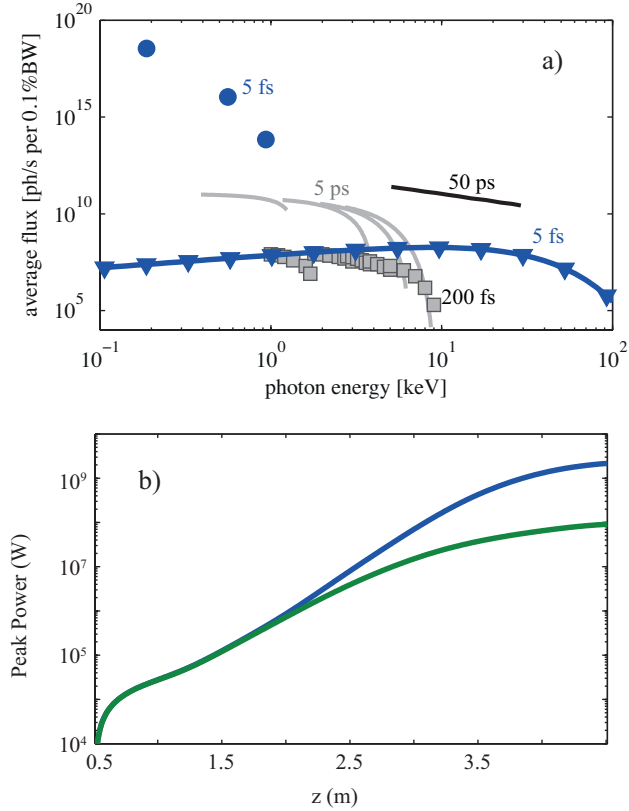}
\caption{
a) Average flux (photons per second per 0.1\% BW) of radiation generated by a \unit[10]{kHz} MP-LWFA compared to existing short pulse systems on 3rd generation light sources.   
Blue circles: \unit[5]{fs}, \unit[10]{kHz}, narrow-band pulses generated by a MP-LWFA-driven FEL mode using a transverse gradient undulator. Blue diamonds:  \unit[5]{fs}, \unit[10]{kHz} broadband betatron pulses generated by a MP-LWFA.  Black line:  \unit[50]{ps}, \unit[0.9]{kHz} broadband pulses generated on the ID9 beam line at ESRF.\cite{Rousse:2007} Grey curves:  \unit[5]{ps} pulses generated at \unit[530]{kHz} at Diamond operating in low alpha mode.\cite{Martin:2011} Grey squares: \unit[200]{fs}, \unit[40]{kHz} pulse generated on a slicing undulator at LLNL.\cite{Rousse:2007}
\newline
b) Peak power of FEL radiation as a function of undulator length generated by a MP-LWFA beam in a standard (solid, green) and TGU (solid, blue) undulator.}
\label{Fig:Radiation}
\end{figure}

Figure \ref{Fig:Radiation}a) shows the calculated average flux of a MP-LWFA-driven betatron source operating at $f_\mathrm{rep} = \unit[10]{kHz}$. For these calculations the betatron oscillation amplitude was assumed to be $r_\beta = \unit[10]{\mu m}$ and the laser focal spot size was decreased to $w_0 = \unit[30]{\mu m}$ to ensure a stronger wake, which was calculated by the fluid code. At a photon energy of \unit[10]{keV} we find that the \emph{average} flux  is $\approx \unit[2 \times 10^{8}]{photons\,s^{-1},per\,0.1 \%\,bandwidth}$.  This average flux is greater than existing beam lines on 3rd generation synchrotron light sources used for sub-picoseond time resolved studies, and with a significantly improved temporal resolution (\unit[5]{fs} compared with $\sim \unit[1]{ps}$). We note that using a still smaller laser spot size would increase the amplitude of the betatron motion, and hence the output photon flux; optimizing the laser-plasma parameters for betatron emission will be the focus of further study.

\subsection{Radiation from undulators}
The MP-LWFA beam can be used to drive spontaneous undulator radiation  with high average brightness and ultra short pulse duration.  More interestingly, the MP-LWFA could be coupled with advanced accelerator transport techniques and undulator technology which are able to accommodate the relatively large energy spread generated by LWFAs; as suggested recently, these techniques include transverse gradient undulators (TGU) or bunch decompression.\cite{Huang:2012, Maier:2012}

In a TGU the magnetic field of the undulator varies with transverse position. If the beam transport system is arranged so that each energy component of the electron beam enters the undulator at a different transverse point then it is possible to match the FEL resonance condition over the entire transverse dimension (and hence energy spectrum) of the electron beam by matching the dispersion function at the entrance of the undulator to the TGU field gradient.

Figure \ref{Fig:Radiation}(b) shows the results of a full 3D time-dependent simulation with the code\cite{Reiche:1999} GENESIS of the peak power of the FEL radiation generated by the MP-LWFA beam in a TGU with the undulator and electron beam parameters reported in Table \ref{Tab:Undulator_parameters}. These show that SASE saturation can be reached in the soft X-ray range ($\lambda_\mathrm{FEL} = \unit[6.9]{nm}$) within a \unit[4]{m} long undulator, yielding a peak power exceeding \unit[1]{GW}. This is a factor 20 larger than the power obtained from a planar undulator with the same electron beam. Harmonics of $\lambda_\mathrm{FEL}$ can also be generated, albeit at significantly reduced power.

The proposed FEL scheme holds the promise of generating femtosecond X-ray pulses of GW peak power at kHz pulse repetition rates, driven by --- and therefore synchronised to --- a compact femtosecond visible laser system.

\begin{table}[tb]
\begin{tabular}{ll}
\hline
Parameter & Value\\
\hline
beam energy & \unit[750]{MeV}\\
rms energy spread & 1\% \\
beam charge & \unit[50]{pC}\\
norm. transverse emittance & $\unit[0.1]{\mu m}$\\
peak current & \unit[10]{kA}\\
bunch duration & \unit[5]{fs} (FWHM)\\
bunch repetition rate & \unit[10]{kHz}\\
\\
undulator type & Superconducting TGU\\
undulator period & \unit[10]{mm}\\
undulator length & \unit[4]{m}\\
undulator parameter	 $K$ & 2.0\\
Transverse gradient & $\unit[150]{m^{-1}}$\\
horizontal dispersion & \unit[1]{cm}\\
resonant wavelength & \unit[6.8]{nm}\\
\hline
\end{tabular}
\caption{Table of the parameters of the electron beam and undulator used in the calculations of FEL radiation driven by a MP-LWFA}
\label{Tab:Undulator_parameters}
\end{table}

\section{Discussion and concluding remarks}
The advantages of MP-LWFA --- such as the potential for high-repetition rate operation and the use of small diameter optics whilst avoiding optical damage --- have been highlighted above.   Here we briefly discuss some potential issues arising from driving plasma accelerators with a train of laser pulses, before some concluding remarks.

A wakefield driven by a single (laser or particle) pulse  can be subject to a deleterious transverse, or `hosing', motion when the tail of the driver is not centred transversely in the wake driven by its head.\cite{Huang:2007jg, Sprangle:1994wa, Najmudin:2003, Kaluza:2010bl} Similar behaviour could also occur in a MP-LWFA if there is a transverse offset between pulses in the train.  We have developed an analytical model\cite{Mangles:2014} of hosing in MP-LWFA which shows that, in the absence of a guiding structure, a single off-axis pulse trailing an on-axis pulse will oscillate transversely with an angular frequency $\omega_{h0}$ which depends on the longitudinal separation of the two pulses and the strength of the wakefield driven by the leading pulse. The oscillation is found to be stable if the longitudinal pulse separation is $(1 - \alpha)\lambda_p$ with $0 < \alpha < \alpha_t \equiv [2(N - 1)]^{-1}$, i.e.\ if the pulses are separated by slightly less than $\lambda_p$. Further, the hosing can be stabilized for  $\alpha > \alpha_t$ by channelling the laser pulses in a waveguide with $\omega_{ch} > \sqrt{(\pi/4)} \omega_{h0}$, where  $\omega_{ch}$ is the frequency that an off-axis pulse would oscillate in the waveguide in the absence of a plasma wave. These conclusions have been confirmed by numerical simulations, as will be reported elsewhere.\cite{Mangles:2014}

We envisage operating MP-LWFA in a linear or quasi-linear regime in which the laser pulses have a peak power below the critical power for relativistic self-focusing.\cite{Esarey:2009} This regime has advantages in that the driving laser pulses propagate without self-focusing, and with reduced spectral shifting and self-steepening; however, it would be necessary to guide the laser pulses over the length of the accelerator using a hollow capillary waveguide\cite{Genoud:2011kj} or plasma channel.\cite{Leemans:2006} In this regime the wake amplitude in a MP-LWFA would be below the threshold for self-trapping;\cite{Esarey:2009, Hooker:2013} this brings a significant advantage in terms of zero dark current, but also means that some method for injecting electrons into the wakefield would be necessary. Controlling electron injection will enable the generation of high-quality, stable electron bunches and for these reasons many techniques are being developed for plasma accelerators.\cite{Faure:2006, Geddes:2008, Gonsalves:2011, Bourgeois:2013ez}

In conclusion MP-LWFA offers a route for enabling plasma accelerators operating at multi-kHz pulse repetition rates to be driven by table-top laser systems with high wall-plug efficiency. Novel plasma accelerators of this type would be ideal for driving compact coherent and incoherent radiation sources providing femtosecond THz to X-ray pulses which are intrinsically synchronized to the driving laser system. The temporal resolution achieved by MP-LWFA-driven incoherent light sources would be at least three orders of magnitude better than possible with extant 3rd generation sources, yet with comparable average photon flux. Likewise MP-LWFAs could drive FELs with comparable peak power to operating FELs, but at a significantly higher repetition rates.

Further developments --- such as plasma accelerator staging --- would allow generation of coherent radiation at shorter wavelengths than considered here. In the longer term the MP-LWFA could provide a stageable, efficient, and high-repetition rate structure capable of reaching beam energies of interest for particle colliders.

\section*{References}

\begin{thebibliography}{10}
\expandafter\ifx\csname url\endcsname\relax
  \def\url#1{\texttt{#1}}\fi
\expandafter\ifx\csname urlprefix\endcsname\relax\def\urlprefix{URL }\fi
\providecommand{\bibinfo}[2]{#2}
\providecommand{\eprint}[2][]{\url{#2}}

\bibitem{Emma:2010}
\bibinfo{author}{Emma, P} \emph{et~al.}
\newblock \bibinfo{title}{{First lasing and operation of an Angstrom-wavelength
  free-electron laser}}.
\newblock \emph{\bibinfo{journal}{Nat. Photonics}}
  \textbf{\bibinfo{volume}{4}}, \bibinfo{pages}{641--647}
  (\bibinfo{year}{2010}).

\bibitem{Leemans:2006}
\bibinfo{author}{Leemans, WP} \emph{et~al.}
\newblock \bibinfo{title}{{GeV} electron beams from a centimetre-scale
  accelerator}.
\newblock \emph{\bibinfo{journal}{Nat. Phys.}} \textbf{\bibinfo{volume}{2}},
  \bibinfo{pages}{696--699} (\bibinfo{year}{2006}).

\bibitem{Kneip:2009}
\bibinfo{author}{Kneip, S} \emph{et~al.}
\newblock \bibinfo{title}{{Near-GeV Acceleration of Electrons by a Nonlinear
  Plasma Wave Driven by a Self-Guided Laser Pulse}}.
\newblock \emph{\bibinfo{journal}{Phys. Rev. Lett.}}
  \textbf{\bibinfo{volume}{103}}, \bibinfo{pages}{035002}
  (\bibinfo{year}{2009}).

\bibitem{Wang:2013}
\bibinfo{author}{Wang, X} \emph{et~al.}
\newblock \bibinfo{title}{{Quasi-monoenergetic laser-plasma acceleration of
  electrons to 2\,GeV}}.
\newblock \emph{\bibinfo{journal}{Nat. Commun.}} \textbf{\bibinfo{volume}{4}},
  \bibinfo{pages}{1--9} (\bibinfo{year}{2013}).

\bibitem{Fuchs:2009}
\bibinfo{author}{Fuchs, M} \emph{et~al.}
\newblock \bibinfo{title}{Laser-driven soft-x-ray undulator source}.
\newblock \emph{\bibinfo{journal}{Nat. Phys.}} \textbf{\bibinfo{volume}{5}},
  \bibinfo{pages}{826--829} (\bibinfo{year}{2009}).

\bibitem{Kneip:2010}
\bibinfo{author}{Kneip, S} \emph{et~al.}
\newblock \bibinfo{title}{{Bright spatially coherent synchrotron X-rays from a
  table-top source}}.
\newblock \emph{\bibinfo{journal}{Nat. Phys.}} \textbf{\bibinfo{volume}{6}},
  \bibinfo{pages}{980--983} (\bibinfo{year}{2010}).

\bibitem{Nakajima:1992}
\bibinfo{author}{Nakajima, K}.
\newblock \bibinfo{title}{Plasma-wave resonator for particle-beam
  acceleration}.
\newblock \emph{\bibinfo{journal}{Phys. Rev. A}} \textbf{\bibinfo{volume}{45}},
  \bibinfo{pages}{1149--1156} (\bibinfo{year}{1992}).

\bibitem{Berezhiani:1992vw}
\bibinfo{author}{Berezhiani, VI} \& \bibinfo{author}{Murusidze, IG}.
\newblock \bibinfo{title}{{Interaction of highly relativistic short laser
  pulses with plasmas and nonlinear wake-field generation}}.
\newblock \emph{\bibinfo{journal}{Physica Scripta}}
  \textbf{\bibinfo{volume}{45}}, \bibinfo{pages}{87} (\bibinfo{year}{1992}).

\bibitem{Umstadter:1994}
\bibinfo{author}{Umstadter, D}, \bibinfo{author}{Esarey, E} \&
  \bibinfo{author}{Kim, J}.
\newblock \bibinfo{title}{Nonlinear plasma waves resonantly driven by optimized
  laser pulse trains}.
\newblock \emph{\bibinfo{journal}{Phys. Rev. Lett.}}
  \textbf{\bibinfo{volume}{72}}, \bibinfo{pages}{1224--1227}
  (\bibinfo{year}{1994}).

\bibitem{Johnson:1994}
\bibinfo{author}{Johnson, DA}, \bibinfo{author}{Cairns, RA},
  \bibinfo{author}{Bingham, R} \& \bibinfo{author}{de~Angelis, U}.
\newblock \bibinfo{title}{Plasma wakefield generation by multiple short
  pulses}.
\newblock \emph{\bibinfo{journal}{Physica Scripta}}
  \textbf{\bibinfo{volume}{T52}}, \bibinfo{pages}{77--81}
  (\bibinfo{year}{1994}).

\bibitem{Dalla:1994b}
\bibinfo{author}{Dalla, S} \& \bibinfo{author}{Lontano, M}.
\newblock \bibinfo{title}{Resonant and quasi-resonant excitation of plasma
  waves by means of sequences of laser pulses}.
\newblock \emph{\bibinfo{journal}{Plasma Phys. Control. Fusion}}
  \textbf{\bibinfo{volume}{36}}, \bibinfo{pages}{1987--2002}
  (\bibinfo{year}{1994}).

\bibitem{Bonnaud:1994uo}
\bibinfo{author}{Bonnaud, G}, \bibinfo{author}{Teychenn{\'e}, D} \&
  \bibinfo{author}{Bobin, JL}.
\newblock \bibinfo{title}{{Wake-field effect induced by laser multiple
  pulses}}.
\newblock \emph{\bibinfo{journal}{Physical Review E}}
  \textbf{\bibinfo{volume}{50}}, \bibinfo{pages}{R36} (\bibinfo{year}{1994}).

\bibitem{Umstadter:1995}
\bibinfo{author}{Umstadter, D}, \bibinfo{author}{Kim, J},
  \bibinfo{author}{Esarey, E}, \bibinfo{author}{Dodd, E} \&
  \bibinfo{author}{Neubert, T}.
\newblock \bibinfo{title}{Resonantly laser-driven plasma waves for electron
  acceleration}.
\newblock \emph{\bibinfo{journal}{Phys. Rev. E}} \textbf{\bibinfo{volume}{51}},
  \bibinfo{pages}{3484--3497} (\bibinfo{year}{1995}).

\bibitem{Kalinnikova:2008}
\bibinfo{author}{Kalinnikova, EI} \& \bibinfo{author}{Levchenko, VD}.
\newblock \bibinfo{title}{{Simulation of the excitation of quasi-plane wake
  waves in a plasma by a resonant sequence of laser pulses with a variable
  envelope}}.
\newblock \emph{\bibinfo{journal}{Plasma Physics Reports}}
  \textbf{\bibinfo{volume}{34}}, \bibinfo{pages}{290--295}
  (\bibinfo{year}{2008}).

\bibitem{Kallos:2008}
\bibinfo{author}{Kallos, E}, \bibinfo{author}{Muggli, P},
  \bibinfo{author}{Katsouleas, T}, \bibinfo{author}{Yakimenko, V} \&
  \bibinfo{author}{Park, J}.
\newblock \bibinfo{title}{Simulations of a high-transformer-ratio plasma
  wakefield accelerator using multiple electron bunches}.
\newblock \emph{\bibinfo{journal}{Advanced Accelerator Concepts: 13th
  Workshop}} \textbf{\bibinfo{volume}{1086}}, \bibinfo{pages}{580--585}
  (\bibinfo{year}{2008}).

\bibitem{Caldwell:2011}
\bibinfo{author}{Caldwell, A} \& \bibinfo{author}{Lotov, KV}.
\newblock \bibinfo{title}{Plasma wakefield acceleration with a modulated proton
  bunch}.
\newblock \emph{\bibinfo{journal}{Phys. Plasmas}}
  \textbf{\bibinfo{volume}{18}}, \bibinfo{pages}{103101}
  (\bibinfo{year}{2011}).

\bibitem{Caldwell:2008}
\bibinfo{author}{Caldwell, A} \& \bibinfo{author}{Lotov, KV}.
\newblock \bibinfo{title}{Experimental results of a plasma wakefield
  accelerator using multiple electron bunches}.
\newblock \emph{\bibinfo{journal}{Proceedings of the 2008 IEEE European
  Particle Accelerator Conference}} \bibinfo{pages}{1912--1914}
  (\bibinfo{year}{2008}).

\bibitem{Corner:2012}
\bibinfo{author}{Corner, L} \emph{et~al.}
\newblock \bibinfo{title}{Multiple pulse resonantly enhanced laser plasma
  wakefield acceleration}.
\newblock \emph{\bibinfo{journal}{AIP Conf. Proc.}}
  \textbf{\bibinfo{volume}{1507}}, \bibinfo{pages}{872--873}
  (\bibinfo{year}{2012}).

\bibitem{Esarey:2009}
\bibinfo{author}{Esarey, E}, \bibinfo{author}{Schroeder, CB} \&
  \bibinfo{author}{Leemans, WP}.
\newblock \bibinfo{title}{Physics of laser-driven plasma-based electron
  accelerators}.
\newblock \emph{\bibinfo{journal}{Rev. Mod. Phys.}}
  \textbf{\bibinfo{volume}{81}}, \bibinfo{pages}{1229--1285}
  (\bibinfo{year}{2009}).

\bibitem{Hooker:2013}
\bibinfo{author}{Hooker, SM}.
\newblock \bibinfo{title}{{Developments in laser-driven plasma accelerators}}.
\newblock \emph{\bibinfo{journal}{Nature Photonics}} \bibinfo{pages}{1--8}
  (\bibinfo{year}{2013}).

\bibitem{Schroeder:2010}
\bibinfo{author}{Schroeder, CB}, \bibinfo{author}{Esarey, E},
  \bibinfo{author}{Geddes, CGR}, \bibinfo{author}{Benedetti, C} \&
  \bibinfo{author}{Leemans, WP}.
\newblock \bibinfo{title}{{Physics considerations for laser-plasma linear
  colliders}}.
\newblock \emph{\bibinfo{journal}{Phys.Rev.ST Accel.Beams}}
  \textbf{\bibinfo{volume}{13}}, \bibinfo{pages}{101301}
  (\bibinfo{year}{2010}).

\bibitem{Lindberg:2006gz}
\bibinfo{author}{Lindberg, RR}, \bibinfo{author}{Charman, AE},
  \bibinfo{author}{Wurtele, JS}, \bibinfo{author}{Friedland, L} \&
  \bibinfo{author}{Shadwick, BA}.
\newblock \bibinfo{title}{{Autoresonant beat-wave generation}}.
\newblock \emph{\bibinfo{journal}{Physics Of Plasmas}}
  \textbf{\bibinfo{volume}{13}}, \bibinfo{pages}{123103}
  (\bibinfo{year}{2006}).

\bibitem{Bulanov:1998}
\bibinfo{author}{Bulanov, S}, \bibinfo{author}{Naumova, N},
  \bibinfo{author}{Pegoraro, F} \& \bibinfo{author}{Sakai, J}.
\newblock \bibinfo{title}{Particle injection into the wave acceleration phase
  due to nonlinear wake wave breaking}.
\newblock \emph{\bibinfo{journal}{Phys. Rev. E}} \textbf{\bibinfo{volume}{58}},
  \bibinfo{pages}{R5257--R5260} (\bibinfo{year}{1998}).

\bibitem{Sprangle:2001}
\bibinfo{author}{Sprangle, P} \emph{et~al.}
\newblock \bibinfo{title}{Wakefield generation and gev acceleration in tapered
  plasma channels}.
\newblock \emph{\bibinfo{journal}{Phys. Rev. E}} \textbf{\bibinfo{volume}{63}},
  \bibinfo{pages}{056405} (\bibinfo{year}{2001}).

\bibitem{Rittershofer:2010}
\bibinfo{author}{Rittershofer, W}, \bibinfo{author}{Schroeder, CB},
  \bibinfo{author}{Esarey, E}, \bibinfo{author}{Gr{\"u}ner, FJ} \&
  \bibinfo{author}{Leemans, WP}.
\newblock \bibinfo{title}{{Tapered plasma channels to phase-lock accelerating
  and focusing forces in laser-plasma accelerators}}.
\newblock \emph{\bibinfo{journal}{Phys. Plasmas}}
  \textbf{\bibinfo{volume}{17}}, \bibinfo{pages}{063104}
  (\bibinfo{year}{2010}).

\bibitem{Miano:1990wf}
\bibinfo{author}{Miano, G}.
\newblock \bibinfo{title}{{Three dimensional analysis of nonlinear plasma
  oscillation}}.
\newblock \emph{\bibinfo{journal}{Physica Scripta}}
  \textbf{\bibinfo{volume}{1990}}, \bibinfo{pages}{198} (\bibinfo{year}{1990}).

\bibitem{OSIRIS}
\bibinfo{author}{Fonseca, R} \emph{et~al.}
\newblock \bibinfo{title}{Osiris: a three-dimensional, fully relativistic
  particle in cell code for modeling plasma based accelerators}.
\newblock \emph{\bibinfo{journal}{Computational Science—ICCS 2002}}
  \bibinfo{pages}{342--351} (\bibinfo{year}{2002}).

\bibitem{Tunnermann:2010}
\bibinfo{author}{T\"{u}nnermann, A}, \bibinfo{author}{Schreiber, T} \&
  \bibinfo{author}{Limpert, J}.
\newblock \bibinfo{title}{Fiber lasers and amplifiers: an ultrafast performance
  evolution}.
\newblock \emph{\bibinfo{journal}{Appl. Opt.}} \textbf{\bibinfo{volume}{49}},
  \bibinfo{pages}{F71--F78} (\bibinfo{year}{2010}).

\bibitem{Morou:2013}
\bibinfo{author}{Mourou, G}, \bibinfo{author}{Brocklesby, B},
  \bibinfo{author}{Tajima, T} \& \bibinfo{author}{Limpert, J}.
\newblock \bibinfo{title}{The future is fibre accelerators}.
\newblock \emph{\bibinfo{journal}{Nat. Photonics}}
  \textbf{\bibinfo{volume}{7}}, \bibinfo{pages}{258--261}
  (\bibinfo{year}{2013}).

\bibitem{Klenke:2013}
\bibinfo{author}{Klenke, A} \emph{et~al.}
\newblock \bibinfo{title}{{530\,W, 1.3\, mJ}, four-channel coherently combined
  femtosecond fiber chirped-pulse amplification system}.
\newblock \emph{\bibinfo{journal}{Opt. Lett.}} \textbf{\bibinfo{volume}{38}},
  \bibinfo{pages}{2283--2285} (\bibinfo{year}{2013}).

\bibitem{Zhou:2007}
\bibinfo{author}{Zhou, S}, \bibinfo{author}{Wise, FW} \&
  \bibinfo{author}{Ouzounov, DG}.
\newblock \bibinfo{title}{Divided-pulse amplification of ultrashort pulses}.
\newblock \emph{\bibinfo{journal}{Opt. Lett.}} \textbf{\bibinfo{volume}{32}},
  \bibinfo{pages}{871--873} (\bibinfo{year}{2007}).

\bibitem{Siders:1998}
\bibinfo{author}{Siders, CW}, \bibinfo{author}{Siders, JLW},
  \bibinfo{author}{Taylor, AJ}, \bibinfo{author}{Park, SG} \&
  \bibinfo{author}{Weiner, AM}.
\newblock \bibinfo{title}{Efficient high-energy pulse-train generation using a
  2 n-pulse michelson interferometer}.
\newblock \emph{\bibinfo{journal}{Appl. Opt.}} \textbf{\bibinfo{volume}{37}},
  \bibinfo{pages}{5302--5305} (\bibinfo{year}{1998}).

\bibitem{Rosenbluth:1972}
\bibinfo{author}{Rosenbluth, MN} \& \bibinfo{author}{Liu, CS}.
\newblock \bibinfo{title}{Excitation of plasma waves by two laser beams}.
\newblock \emph{\bibinfo{journal}{Phys. Rev. Lett.}}
  \textbf{\bibinfo{volume}{29}}, \bibinfo{pages}{701--705}
  (\bibinfo{year}{1972}).

\bibitem{Amiranoff:1992}
\bibinfo{author}{Amiranoff, F} \emph{et~al.}
\newblock \bibinfo{title}{Observation of modulational instability in nd-laser
  beat-wave experiments}.
\newblock \emph{\bibinfo{journal}{Phys. Rev. Lett.}}
  \textbf{\bibinfo{volume}{68}}, \bibinfo{pages}{3710--3713}
  (\bibinfo{year}{1992}).

\bibitem{Weiner:1990}
\bibinfo{author}{Weiner, A} \& \bibinfo{author}{Leaird, D}.
\newblock \bibinfo{title}{{Generation of terahertz trains of femtosecond pulses
  by phase-only filtering}}.
\newblock \emph{\bibinfo{journal}{Opt. Lett.}} \textbf{\bibinfo{volume}{15}},
  \bibinfo{pages}{51} (\bibinfo{year}{1990}).

\bibitem{Deutsch:1991}
\bibinfo{author}{Deutsch, M}, \bibinfo{author}{Meerson, B} \&
  \bibinfo{author}{Golub, JE}.
\newblock \bibinfo{title}{Strong plasma wave excitation by a ``chirped'' laser
  beat wave}.
\newblock \emph{\bibinfo{journal}{Phys. Fluids B}}
  \textbf{\bibinfo{volume}{3}}, \bibinfo{pages}{1773--1780}
  (\bibinfo{year}{1991}).

\bibitem{Phuoc:2012cg}
\bibinfo{author}{Phuoc, KT} \emph{et~al.}
\newblock \bibinfo{title}{{All-optical Compton gamma-ray source}}.
\newblock \emph{\bibinfo{journal}{Nature Photonics}} \bibinfo{pages}{1--4}
  (\bibinfo{year}{2012}).

\bibitem{Powers:2013bx}
\bibinfo{author}{Powers, ND} \emph{et~al.}
\newblock \bibinfo{title}{{Quasi-monoenergetic and tunable X-raysfrom a
  laser-driven Compton light source}}.
\newblock \emph{\bibinfo{journal}{Nature Photonics}} \bibinfo{pages}{1--4}
  (\bibinfo{year}{2013}).

\bibitem{Esarey:2002hk}
\bibinfo{author}{Esarey, E}, \bibinfo{author}{Shadwick, B},
  \bibinfo{author}{Catravas, P} \& \bibinfo{author}{Leemans, W}.
\newblock \bibinfo{title}{{Synchrotron radiation from electron beams in
  plasma-focusing channels}}.
\newblock \emph{\bibinfo{journal}{Physical Review E}}
  \textbf{\bibinfo{volume}{65}}, \bibinfo{pages}{056505}
  (\bibinfo{year}{2002}).

\bibitem{Rousse:2007}
\bibinfo{author}{Rousse, A}, \bibinfo{author}{Phuoc, KT},
  \bibinfo{author}{Shah, R}, \bibinfo{author}{Fitour, R} \&
  \bibinfo{author}{Albert, F}.
\newblock \bibinfo{title}{{Scaling of betatron X-ray radiation}}.
\newblock \emph{\bibinfo{journal}{The European Physical Journal D}}
  \textbf{\bibinfo{volume}{45}}, \bibinfo{pages}{391--398}
  (\bibinfo{year}{2007}).

\bibitem{Martin:2011}
\bibinfo{author}{Martin, IPS}, \bibinfo{author}{Rehm, G},
  \bibinfo{author}{Thomas, C} \& \bibinfo{author}{Bartolini, R}.
\newblock \bibinfo{title}{Experience with low-alpha lattices at the diamond
  light source}.
\newblock \emph{\bibinfo{journal}{Phys. Rev. ST Accel. Beams}}
  \textbf{\bibinfo{volume}{14}}, \bibinfo{pages}{040705}
  (\bibinfo{year}{2011}).

\bibitem{Huang:2012}
\bibinfo{author}{Huang, Z}, \bibinfo{author}{Ding, Y} \&
  \bibinfo{author}{Schroeder, CB}.
\newblock \bibinfo{title}{{Compact X-ray Free-Electron Laser from a
  Laser-Plasma Accelerator Using a Transverse-Gradient Undulator}}.
\newblock \emph{\bibinfo{journal}{Phys. Rev. Lett.}}
  \textbf{\bibinfo{volume}{109}}, \bibinfo{pages}{204801}
  (\bibinfo{year}{2012}).

\bibitem{Maier:2012}
\bibinfo{author}{Maier, A} \emph{et~al.}
\newblock \bibinfo{title}{{Demonstration Scheme for a Laser-Plasma-Driven
  Free-Electron Laser}}.
\newblock \emph{\bibinfo{journal}{Phys. Rev. X}} \textbf{\bibinfo{volume}{2}},
  \bibinfo{pages}{031019} (\bibinfo{year}{2012}).

\bibitem{Reiche:1999}
\bibinfo{author}{Reiche, S}.
\newblock \bibinfo{title}{{GENESIS} 1.3: a fully 3d time-dependent {FEL}
  simulation code}.
\newblock \emph{\bibinfo{journal}{Nucl. Inst. Meth. Res. A}}
  \textbf{\bibinfo{volume}{429}}, \bibinfo{pages}{243--248}
  (\bibinfo{year}{1999}).

\bibitem{Huang:2007jg}
\bibinfo{author}{Huang, C} \emph{et~al.}
\newblock \bibinfo{title}{{Hosing Instability in the Blow-Out Regime for
  Plasma-Wakefield Acceleration}}.
\newblock \emph{\bibinfo{journal}{Physical Review Letters}}
  \textbf{\bibinfo{volume}{99}}, \bibinfo{pages}{255001}
  (\bibinfo{year}{2007}).

\bibitem{Sprangle:1994wa}
\bibinfo{author}{Sprangle, P}, \bibinfo{author}{Krall, J} \&
  \bibinfo{author}{Esarey, E}.
\newblock \bibinfo{title}{{Hose-modulation instability of laser pulses in
  plasmas}}.
\newblock \emph{\bibinfo{journal}{Physical Review Letters}}
  \textbf{\bibinfo{volume}{73}}, \bibinfo{pages}{3544--3547}
  (\bibinfo{year}{1994}).

\bibitem{Najmudin:2003}
\bibinfo{author}{Najmudin, Z} \emph{et~al.}
\newblock \bibinfo{title}{Self-modulated wakefield and forced laser wakefield
  acceleration of electrons}.
\newblock \emph{\bibinfo{journal}{Phys. Plasmas}}
  \textbf{\bibinfo{volume}{10}}, \bibinfo{pages}{2071--2077}
  (\bibinfo{year}{2003}).

\bibitem{Kaluza:2010bl}
\bibinfo{author}{Kaluza, M} \emph{et~al.}
\newblock \bibinfo{title}{{Measurement of Magnetic-Field Structures in a
  Laser-Wakefield Accelerator}}.
\newblock \emph{\bibinfo{journal}{Physical Review Letters}}
  \textbf{\bibinfo{volume}{105}}, \bibinfo{pages}{115002}
  (\bibinfo{year}{2010}).

\bibitem{Mangles:2014}
\bibinfo{author}{Mangles, S}, \bibinfo{author}{Cole, JM},
  \bibinfo{author}{Corner, L}, \bibinfo{author}{Walczak, R} \&
  \bibinfo{author}{Hooker, SM}.
\newblock \bibinfo{title}{Analysis of hosing in a multi-pulse laser wakefield
  accelerator}.
\newblock \emph{\bibinfo{journal}{To be submitted}}  (\bibinfo{year}{2014}).

\bibitem{Genoud:2011kj}
\bibinfo{author}{Genoud, G} \emph{et~al.}
\newblock \bibinfo{title}{{Laser-plasma electron acceleration in dielectric
  capillary tubes}}.
\newblock \emph{\bibinfo{journal}{Applied Physics B}} \bibinfo{pages}{1--8}
  (\bibinfo{year}{2011}).

\bibitem{Faure:2006}
\bibinfo{author}{Faure, J} \emph{et~al.}
\newblock \bibinfo{title}{Controlled injection and acceleration of electrons in
  plasma wakefields by colliding laser pulses}.
\newblock \emph{\bibinfo{journal}{Nature}} \textbf{\bibinfo{volume}{444}},
  \bibinfo{pages}{737--739} (\bibinfo{year}{2006}).

\bibitem{Geddes:2008}
\bibinfo{author}{Geddes, CGR} \emph{et~al.}
\newblock \bibinfo{title}{Plasma-density-gradient injection of low
  absolute-momentum-spread electron bunches}.
\newblock \emph{\bibinfo{journal}{Phys. Rev. Lett.}}
  \textbf{\bibinfo{volume}{100}}, \bibinfo{pages}{215004}
  (\bibinfo{year}{2008}).

\bibitem{Gonsalves:2011}
\bibinfo{author}{Gonsalves, AJ} \emph{et~al.}
\newblock \bibinfo{title}{{Tunable laser plasma accelerator based on
  longitudinal density tailoring}}.
\newblock \emph{\bibinfo{journal}{Nat. Phys.}} \textbf{\bibinfo{volume}{7}},
  \bibinfo{pages}{862--866} (\bibinfo{year}{2011}).

\bibitem{Bourgeois:2013ez}
\bibinfo{author}{Bourgeois, N}, \bibinfo{author}{Cowley, J} \&
  \bibinfo{author}{Hooker, SM}.
\newblock \bibinfo{title}{{Two-Pulse Ionization Injection into Quasilinear
  Laser Wakefields}}.
\newblock \emph{\bibinfo{journal}{Physical Review Letters}}
  \textbf{\bibinfo{volume}{111}}, \bibinfo{pages}{155004}
  (\bibinfo{year}{2013}).

\end{thebibliography}


\section{Acknowledgments}
This work was supported by the Engineering and Physical Sciences Research Council [grant no.\ EP/H011145/1], the Leverhulme Trust [grant no.\ F/08 776/G],  the Science and Technology Facilities Council [grant no. \ ST/J002011] and the Royal Society.

\pagebreak
\section{Methods}
The radiation generated by betatron motion within the MP-LWFA was estimated using standard theory,\cite{Esarey:2002hk} noting that the MP-LWFA will not operate in the ``blow-out'' regime and that therefore the betatron oscillation wavelength is given by $\lambda_\beta = \pi w_0 \sqrt{ 2\gamma /\hat{\Phi}_0}$ where ${\hat{\Phi}}_0 = e\Phi_0/(m_ec^2)$ is the normalised maximum potential of the plasma wave. The number of photons emitted can be estimated to be$ N_{ph} = (2\pi/3) r_e m_e c^2 \gamma^2 k_\beta K^2 N_e N_\beta / E_c$
where $k_\beta = 2\pi/\lambda_\beta$ is the betatron wavenumber; $K = \gamma k_\beta r_\beta$ is the wiggler strength parameter for oscillation of amplitude $r_\beta$;  $N_e$ is the number of electrons in the bunch; $N_\beta$ is the number of betatron oscillations and $E_c = \frac{3}{2}\hbar\gamma^2 K k_\beta c$ is the critical energy of the synchrotron-like spectrum of the radiation.  

\end{document}